\documentclass[aps,prl,groupedaddress,twocolumn,showpacs,amsfonts,amsmath,amssymb]{revtex4}
\usepackage{graphicx}
\usepackage{bm}

\begin{document}

\title{Theory of Upper Critical Field without Energy Quantization}
\author{L. Wang} 
\altaffiliation[Present address:]{
Spintronics, Media and Interface Division, 
Data Storage Institute, Singapore 117608}
\email{wang_li@dsi.a-star.edu.sg}
\author{H. S. Lim}
\author{C. K. Ong}
\affiliation{
Department of Physics, National University of Singapore, Singapore 117542 
}

\begin{abstract}
{
Conventional theories for determining upper critical fields are inevitably 
related to the lowest eigenvalues of appropriate equations. 
In this Letter, a new theory of upper critical fields is designed and justified. 
Using MgB$_2$ as modeling prototype, our computations are 
in excellent agreement with the Ginzburg-Landau theory. 
The long-standing issue, the upward  curvature of the upper critical field, 
is found to be a manifestation of the crossover of the order parameter. 
The current theory is an alternative to the traditional technique of 
energy quantization in determining upper critical fields. 
}
\end{abstract}

\maketitle

The upper critical field ($B_{c2}$) is the maximum magnetic field that a 
bulk type-II superconductor \cite{Abrikosov57} can sustain in the superconducting state. 
The value of $B_{c2}$ is very important as it partially determines 
the current carrying capacity of the superconductor and 
its uses e.g. to produce high field superconducting magnets. 
Furthermore, study of properties of $B_{c2}$ may test the validity of 
various theoretical models and provide information for 
important superconducting parameters such as the coherence length. 
Hence, research of the upper critical field is of practical, fundamental and 
enduring interest. 

To determine the upper critical field, an important starting point is 
the linearized Ginzburg-Landau (GL) equation \cite{Ginzburg50} and 
its variants. Examples would be 
the quantum harmonic oscillator equation \cite{Abrikosov57}, 
the Mathieu equation \cite{Klemm75}, 
the $p$-wave GL equation \cite{Zhu97}, and 
the $d$-wave GL equation \cite{Chang98}. 
Determining upper critical fields from these equations involves finding 
the lowest eigenvalues of the said equations. Alternatively, the microscopic 
description of the upper critical field, based on the Gor'kov gap 
equation \cite{Gorkov59Bc2}, may also be reduced to 
finding the lowest eigenvalues of appropriate 
equations \cite{Helfand64,Klemm81,Takahashi86,Galitski03}. 
Note that the eigenstate \cite{Abrikosov57} of the lowest eigenvalue of 
the linearized GL equation is usually chosen as the trial function in 
the variational determinations of $B_{c2}$ (see, e.g., 
Refs.~\cite{Gorkov59Bc2,Berlinsky95}). Thus, the variational treatment 
is related to the lowest-eigenvalue method. Note further that 
the perturbation approach to determining upper critical fields 
is also intimately related to the scenario of 
the lowest eigenvalue \cite{Joynt90}. Consequently, we are led to 
conclude that almost all the efforts 
\cite{Abrikosov57,Ginzburg50,Klemm75,Zhu97,Chang98,Gorkov59Bc2,Berlinsky95,
Helfand64,Klemm81,Takahashi86,Galitski03,Joynt90} for determining 
upper critical fields are related to the lowest eigenvalue of 
an appropriate equation, which is in turn, more or less, explicitly or 
implicitly, related to the important concept of energy quantization 
\cite{Abrikosov57,Ginzburg50,Planck00,Heisenberg25,Schrodinger26}. 
This traditional framework of determining upper critical fields 
by energy-quantization related techniques or scenarios has existed 
for fifty years or so \cite{Abrikosov57,Ginzburg50,Gorkov59Bc2}. 
However, in this Letter, a new theory will be reported and justified 
to determine upper critical fields without recourse to 
energy quantization. Some applications and implications of 
the theory proposed will also be addressed.  

Various superconductors are layered compounds although they may be 
oxides, metallic or organic. To describe layered superconductors, 
a continuous Ginzburg-Landau (CGL) model was proposed \cite{Koyama92}, 
in which the GL coefficients and the superpair masses are 
assumed spatially dependent. This model was considered \cite{Koyama92} to 
approach the limiting cases of the anisotropic Ginzburg-Landau (AGL) theory 
(see Ref.~\cite{Clem98} and references therein) and 
the Lawrence-Doniach (LD) model \cite{Lawrence71}.    

Recently, a modified CGL model has been proposed \cite{Wang01a,Wang01b}. 
The unit cell in this model consists of alternating superconducting and 
weakly superconducting layers. The $z$-axis is normal to the layers and 
its origin is at the midpoint of one of the weakly superconducting layers. 
The center of the superconducting layer is located at $D$/2, 
where $D$ is the size of the unit cell \cite{Wang01a,Wang02a}.  
Applying an external magnetic field $B$ parallel to the layers, 
the linearized CGL equation may be written 
as \cite{Koyama92,Wang01a,Wang01b,Wang02a}
\begin{eqnarray}
-\frac{\hbar^{2}}{2M(z)}\frac{\partial^{2}}{\partial z^{2}}\Psi(z) -
\frac{\hbar^{2}}{2}\left[ \frac{\partial}{\partial z} \frac{1}{M(z)}\right] 
\frac{\partial}{\partial z} \Psi(z) + \nonumber \\
\left[ \alpha(T,z) + \frac{1}{2m(z)}(2eB)^{2}(z-\frac{D}{2})^{2} \right] 
\Psi(z) = 0, 
\label{eq:linearCGL}
\end{eqnarray}
where the nucleation center ($z_c=\hbar k_x/2eB$ ) is set at $D$/2. 
The condensation coefficient $\alpha(T,z)$ and the effective masses, 
$M(z)$ and $m(z)$, are assumed as \cite{Wang01a,Wang01b,Wang02a}  
\begin{subequations}\label{eq:coeficient} 
\begin{eqnarray}
\alpha (T,z)   & =&  \left[ \alpha_0 + 
\alpha_{1}\cos (2\pi z/D)\right] (1-T/T_{c}), 
\label{eq:coeficienta} \\ 
\frac{1}{M(z)} & = & G_{0} + G_{1} \cos (2\pi z/D), 
\label{eq:coeficientb} \\
\frac{1}{m(z)} & = & g_{0} + g_{1} \cos (2\pi z/D).
\label{eq:coeficientc}
\end{eqnarray}  
\end{subequations}
Here $\alpha_0$, $\alpha_1$, $G_0$, $G_1$, $g_0$ and $g_1$ are 
model parameters and the determinations of their values can be 
found in Ref.~\cite{Wang01b}. 

To determine the upper critical field, Eq.~(\ref{eq:linearCGL})  
should be completed by appropriate boundary conditions. 
Here we choose the open boundary conditions (OBCs), 
\begin{equation}
\left. \Psi(z)\right|_{z \rightarrow \pm \infty} =0.
\label{eq:OBC}
\end{equation}  
Eqs.~(\ref{eq:linearCGL}) and (\ref{eq:OBC}) can be 
transformed into a system of the form 
\begin{equation}
\bm{U\Psi}=0, 
\label{eq:UPsi}
\end{equation}
where $\bm{\Psi}$ is a wave function representing the discrete solutions 
of Eq.~(\ref{eq:linearCGL}) and $\bm{U}$ is the corresponding 
coefficient matrix \cite{Wang02b}. For Eq.~(\ref{eq:UPsi}) to 
have non-trivial solutions, the determinant of $\bm{U}$ should be zero,
\begin{equation}
\text{det}|\bm{U}|=0. 
\label{eq:det_U}
\end{equation}
It can be verified \cite{Wang02b} that the parameters of 
the magnetic field $B$ appear only in the main diagonal of 
$\bm{U}$ so that we have 
\begin{equation}
\det|\bm{P}-B^{2}\bm{I}|=0, 
\label{eq:det_P}
\end{equation}
where $\bm{I}$ is a unitary matrix and $\bm{P}$ is a sparse matrix 
independent of $B$. Thus, the maximum $B$, namely 
the upper critical field, can be deduced from 
the largest eigenvalue of the following eigen equation
\begin{equation}
\bm{P\chi}=B^{2}\bm{\chi}, 
\label{eq:P_eigen}
\end{equation}
where $\bm{\chi}$ is the eigenfunction of $\bm{P}$ and this function is 
an auxiliary field to Eq.~(\ref{eq:det_P}). Having determined $B_{c2}$, 
one can obtain the corresponding order parameter by substituting $B_{c2}$ 
back into Eq.~(\ref{eq:UPsi}). 

It should be stressed that Eq.~(\ref{eq:linearCGL}) can be written as 
\begin{equation}
\text{{\it \^{Q}}}\Psi(z)=B^{2}\Psi(z),
\label{eq:Q_operator}
\end{equation}
where \text{{\it \^{Q}}} is an operator. Eq.~(\ref{eq:Q_operator}), 
together with the boundary conditions of Eq.~(\ref{eq:OBC}), can be  
written as the following matrix eigen equation \cite{Wang02b}, 
\begin{equation}
\bm{Q\Psi}=B^{2}\bm{\Psi},
\label{eq:Q_eigen}
\end{equation}
where $\bm{\Psi}$ possesses the same meaning as in Eq.~(\ref{eq:UPsi}). 
$\bm{Q}$ is the same as $\bm{P}$ in Eq.~(\ref{eq:P_eigen}). 
Hence, Eq.~(\ref{eq:Q_eigen}) is equivalent to Eq.~(\ref{eq:P_eigen}) 
and $\bm{\Psi}$ plays the same role as $\bm{\chi}$ as eigenfunctions.

\begin{figure}[t]
\includegraphics[bb=106 381 446 665,scale=0.61,clip]{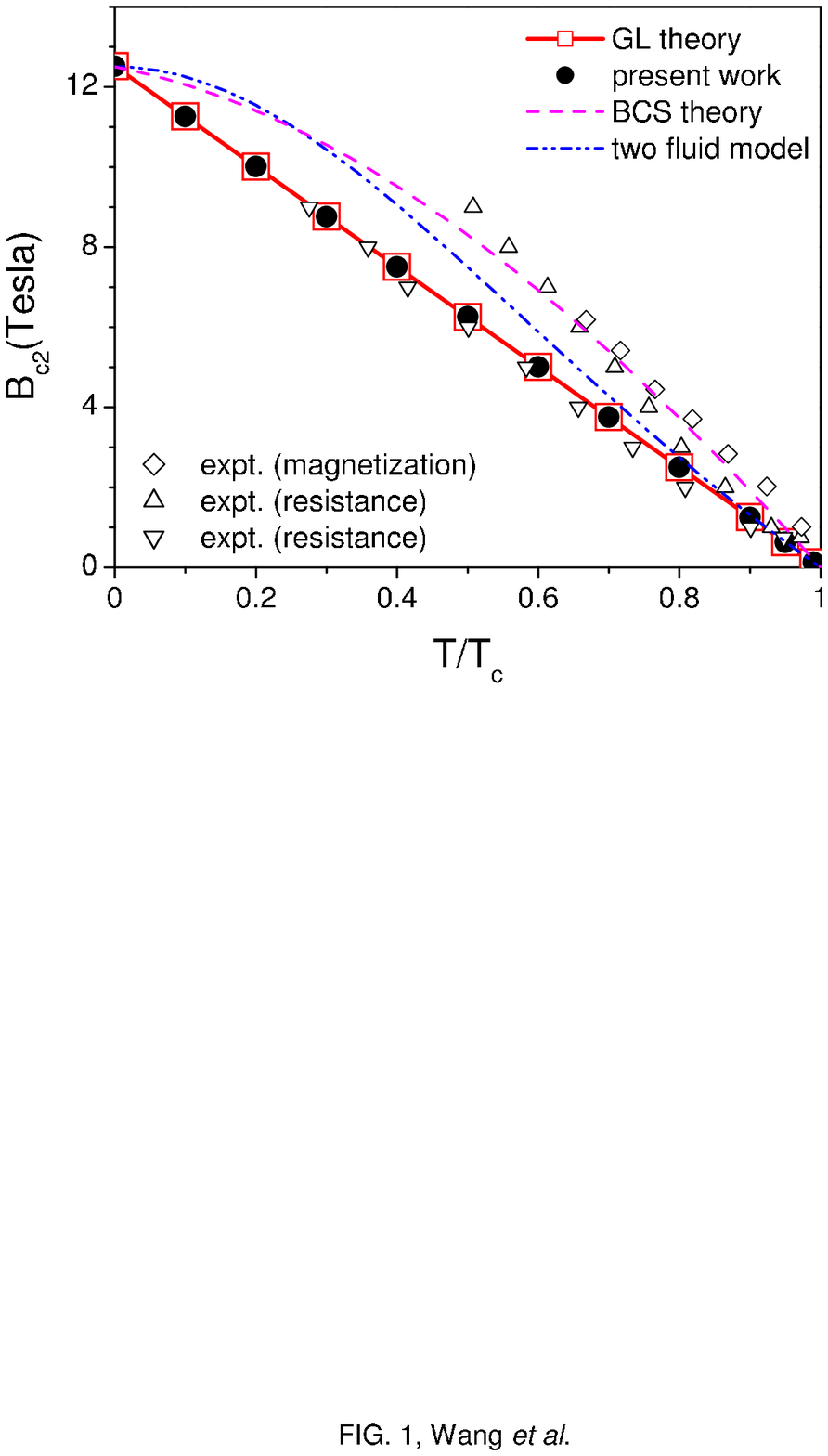} 
\caption{
Temperature dependences of the upper critical field of MgB$_2$. 
The calculated data ({\Large$\bullet$}) are in excellent agreement 
with the traditional GL theory of Eq.~(\ref{eq:tradition_Bc2}) 
({\small$\square$}) and reasonably consistent with the experiments in 
Ref.~\cite{Finnemore01} [data ({\Large $\diamond$}) from the 
magnetization measurements; ($\bigtriangleup$) and ($\bigtriangledown$) 
from the measurements of the resistance-temperature curve near 
the superconducting and normal states, respectively]. 
For comparison, results from the BCS theory and the two fluid 
model \cite{Woollam74} are also illustrated. 
} 
\label{fig:Bc2_T}
\end{figure}

For comparison of our determinations with the traditional treatment for 
bulk superconductors \cite{Abrikosov57,Ginzburg50,Ginzburg57,Tinkham96}, 
we model the MgB$_2$ superconductor \cite{Nagamatsu01} 
whose anisotropy is small \cite{Wang01b,Finnemore01}. For convenience, 
we treat it to be isotropic and ignore the spatial dependences in 
Eq.~(\ref{eq:coeficient}). Hence, $G_0=g_0=1/m=0.5$ and 
$G_1=g_1=\alpha_1=0$ were chosen and the nucleation center 
set at $z_c=0$. Other parameters such as $\alpha_0$ for MgB$_2$ are 
taken from Ref.~\cite{Wang01b}. It was found that 
(see Fig.~\ref{fig:Bc2_T}) the results obtained from 
Eq.~(\ref{eq:Q_eigen}) are strikingly in agreement with 
the traditional GL work \cite{Abrikosov57,Ginzburg50,Ginzburg57,Tinkham96},
\begin{equation}
B_{c2} = \Phi_0/2\pi\xi^2=\sqrt{2}\kappa B_c, 
\label{eq:tradition_Bc2}
\end{equation} 	           			  	    
where the flux quantum $\Phi_0=h/2e$ and the coherence length 
$\xi=\hbar/\sqrt{2m|\alpha(T)|}=\hbar / \sqrt{2|\alpha_0|(1-T/T_c)/g_0}$. 
$B_c$ is the thermal dynamic critical field and $\kappa$ is 
the ratio of the penetration depth $\lambda$ to the coherence length 
$\xi$. It should be emphasized that Eq.~(\ref{eq:tradition_Bc2}) is 
derived from energy quantization of the quantum harmonic oscillator 
equation \cite{Ginzburg50,Abrikosov57,Tinkham96} and the microscopic 
theory \cite{Gorkov59Bc2} may also arrive at a similar expression. 
The agreement of our calculations with Eq.~(\ref{eq:tradition_Bc2}) 
lends us strong credence to our theory of upper critical fields. 
Furthermore, we found that the calculated results are reasonably 
consistent with the experiments of Finnemore 
{\it et al.} \cite{Finnemore01} and with some theoretical 
results \cite{Wang02b,Woollam74}, as shown in Fig.~\ref{fig:Bc2_T}. 
These agreement and consistencies show that MgB$_2$ is a superconductor 
describable within the GL/BCS framework. Note, however, that 
our main interest here lies in the qualitative predictions of 
the theories and thus we have assumed that the theories arrive at 
the same value of $B_{c2}$ at zero temperature; 
otherwise, the actual values are deviated from the 
zero-temperature one but the qualitative trends remained.   

\begin{figure}[pt]
\includegraphics[bb=95 383 458 671,scale=0.61,clip]{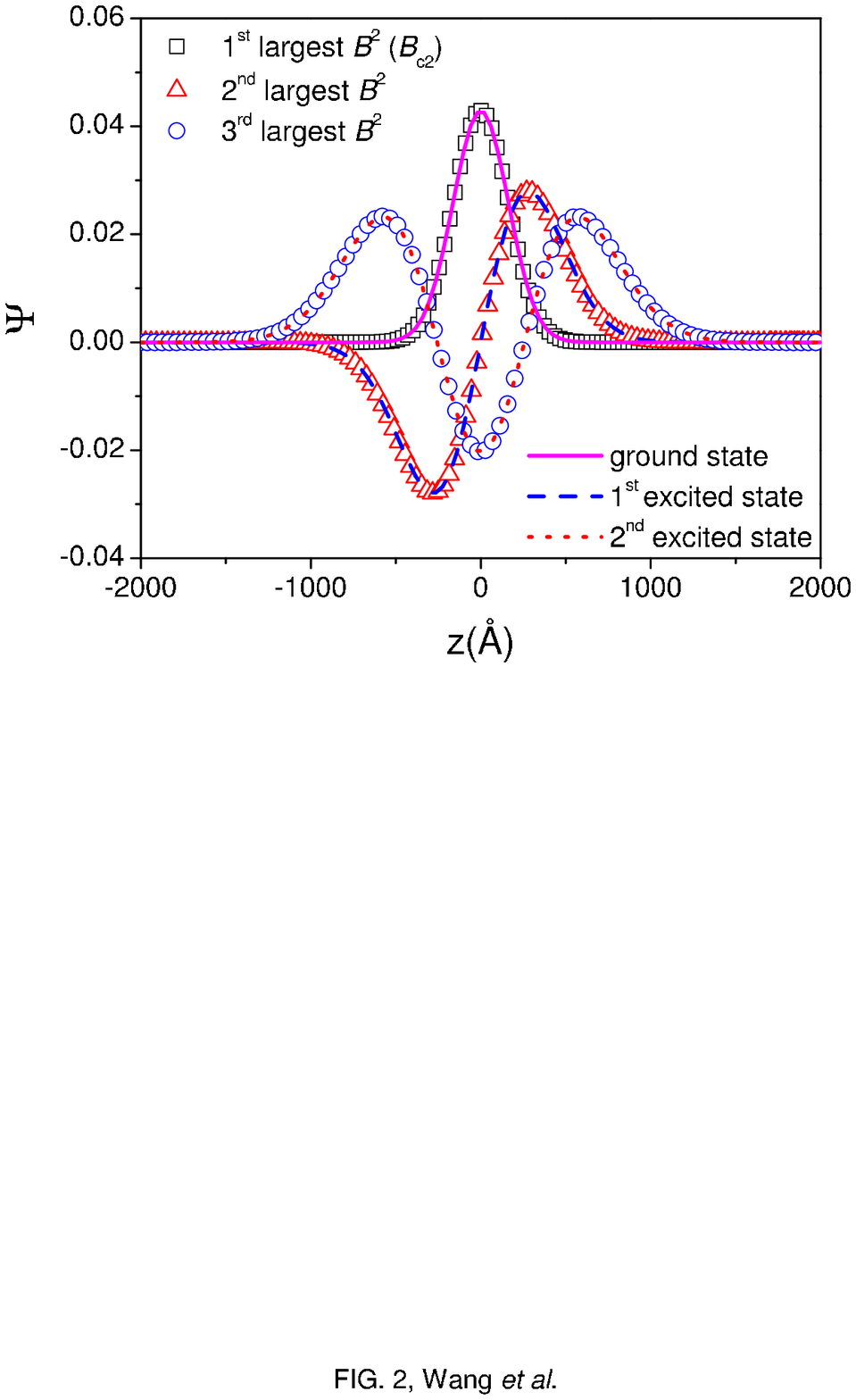} 
\caption{ 
Order parameters at the upper critical field, the second and third 
largest square of the magnetic field respectively correspond to 
the ground, first and second excited states of 
the quantum harmonic oscillator [Eq.~(\ref{eq:wavefunction})]. 
}
\label{fig:wavefunction}
\end{figure}

In Fig.~\ref{fig:wavefunction}, we have plotted the spatial distribution 
of the order parameter at $B_{c2}$, which was obtained from 
Eq.~(\ref{eq:Q_eigen}) at $t=(T/T_c)=0.9$ for bulk MgB$_2$. 
It can be seen that the order parameter is in excellent 
agreement with the following function at $n=0$ (in a.u.)
\begin{equation}
\Psi_n(z) = \frac{1}{\pi^{1/4}\zeta^{1/2}\sqrt{2^n n!}}
\exp\left(-\frac{z^2}{2\zeta^2}\right) H_n\left(\frac{z}{\zeta}\right),
\label{eq:wavefunction}
\end{equation}
where $\zeta=1/\sqrt{2B_{c2}}$ and $H_n$ are Hermite polynomials. 
Eq.~(\ref{eq:wavefunction}) represents the eigenstates of 
the harmonic oscillator equation with energy serving as 
the eigenvalues \cite{Landau77}. The function at $n=0$ is the 
state of the lowest energy eigenvalue. The remarkable agreements of 
our calculations with Eqs.~(\ref{eq:tradition_Bc2}) and 
(\ref{eq:wavefunction}) unambiguously justifies the current theory of 
upper critical fields. 

It is interesting to find that (see Fig.~\ref{fig:wavefunction}) 
the order parameters corresponding to the second and third largest 
$B^2$ are nothing else but the first and second excited states of 
the harmonic oscillator equation, as given by Eq.~(\ref{eq:wavefunction}) 
at $n=1,2$, respectively. Note that in the expression of 
$\zeta=1/\sqrt{2B_{c2}}$, $B_{c2}$ should now be replaced by 
$B_2$ and $B_3$, which are the square roots of the second and third largest 
$B^2$, respectively. Here, a few points are worthy of notice.  
(i) What does $B$ associated with the lowest eigenvalue of $B^2$ mean? 
Which state are superconductors in subject to such $B$? 
The respective possible answers might be the lower critical field and 
the Meissner state? 
(ii) Are there any physical consequences in Eq.~(\ref{eq:Q_eigen}) [or 
Eq.~(\ref{eq:P_eigen})] by changing $2e$ to $e$ in Eq.~(\ref{eq:linearCGL})? 
Are there any other physical significances in Eqs.~(\ref{eq:P_eigen}) and 
(\ref{eq:Q_eigen}) themselves except being used to obtain 
upper critical fields? 
(iii) The sign of $\alpha_0$ in Eq.~(\ref{eq:coeficienta}) may need attention: 
by setting positive condensation energy $\alpha_0$ (with $\alpha_1=0$), 
it was found that the calculated upper critical field is always zero. 
Hence, negative condensation energy is a necessity for 
a superconducting state, as expected. 

\begin{figure}[t]
\includegraphics[bb=105 252 453 707,scale=0.61,clip]{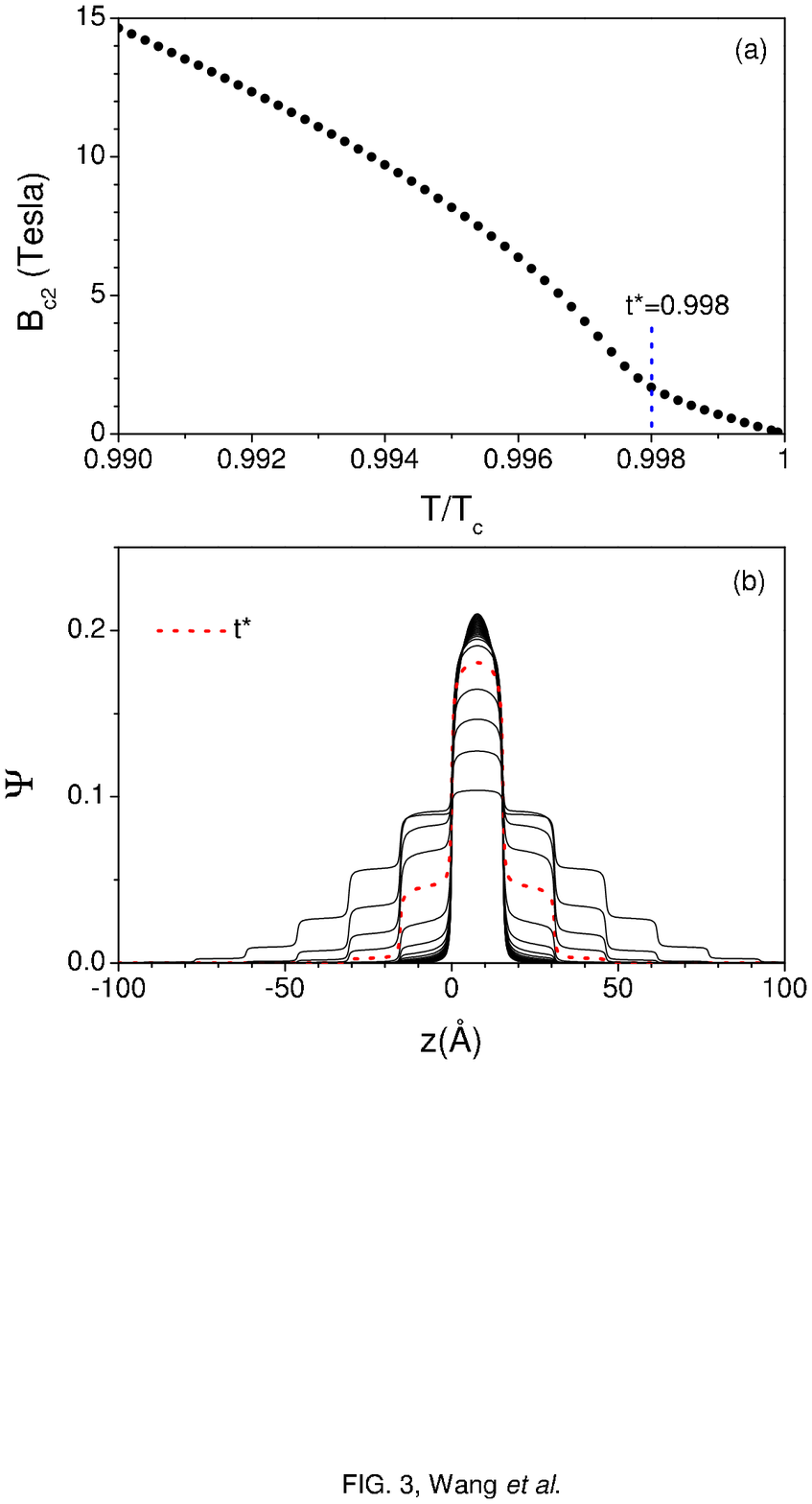} 
\caption{
Upper critical field and order parameter of Bi2212 near $T_c$. 
(a) Upper critical field as a function of reduced temperature $t=T/T_c$, 
showing an upward curvature at $t^* \sim 0.998$. 
(b) Spatial distributions of the order parameters near $T_c$. 
For $t<t^*$, the order parameters are mainly confined 
within one unit cell whereas for $t>t^*$, the order parameters 
spread over several unit cells. This crossover of the order parameter 
manifests itself as the upward curvature observed in (a). 
}
\label{fig:upward_crossover} 
\end{figure}

Now some applications and implications of our theory will be addressed.  
The upward (positive) curvature of the $B_{c2}$-$T$  curve in 
layered superconducting systems is a subject of long-term interest 
\cite{Klemm75,Takahashi86,Joynt90,Lawrence71,Wang02a,
Woollam74,Ruggiero80,Sundaram91,Alexandrov96,Abrikosov97,Shulga98,Brandow98,
Welp03,Domanski03}. In Fig.~\ref{fig:upward_crossover}(a), 
we have plotted the calculated parallel upper critical field as 
a function of the reduced temperature for the highly anisotropic 
cuprate superconductor Bi$_2$Si$_2$CaCu$_2$O$_8$ (Bi2212). 
The interval of the reduced temperature $\Delta t$ for the plot is 0.0002. 
It is clear that at $t^* \sim 0.998$, an upward curvature appears. 
In the extreme vicinity of $T_c$, the $B_{c2}$-$T$ plot is linear 
(which is consistent with the GL and AGL theories). 
A linear behavior also holds at low temperatures \cite{Wang02a,Wang02b}. 
Hence, a transition exists linking the two kinds of linear behaviors. 
In Fig.~\ref{fig:upward_crossover}(b), the spatial distributions of 
the corresponding order parameters are presented, 
where $\Delta t=0.0004$ was chosen for clarity. It is clear that 
the order parameters below $t^* \sim 0.998$ mainly reside within 
one unit cell [i.e., two dimensional (2D) behavior] whereas 
those above $t^*$ spread into the neighboring cells. 
With increasing temperature, more layers are penetrated 
[i.e., three dimensional (3D) behavior]. Thus, $t^*$ is a crossover 
temperature for the order parameter to transit between the 2D and 
3D behaviors. Clearly, the upward curvature 
[Fig.~\ref{fig:upward_crossover}(a)] is originated from the crossover of 
the order parameter [Fig.~\ref{fig:upward_crossover}(b)], 
which is in turn due to that the layering structures 
[cf. Eq.~(\ref{eq:coeficient})] come into effects. 
Note that the value of $t^* \sim 0.998$ obtained here for Bi2212 is 
consistent with Ref.~\cite{Sokolov91}, in which the most likely value 
was estimated in the range 0.997-0.998. 

It is worth mentioning that by considering the small anisotropy in 
MgB$_2$ ($G_0 \sim 1$, $g_0 \sim 1$ and $G_1=g_1=\alpha_1=0$), 
our calculations are consistent with the AGL theory. Moreover, 
CGL simulations have also been performed for MgB$_2$ but 
no obvious upward curvature was observed. In fact, the issue of the 
upward feature is quite complicated \cite{Wang02a,Brandow98,Domanski03}. 
A recent experimental study \cite{Welp03} shows that the 
upward curvature may also exist in MgB$_2$. If the same reason 
for the curvature, i.e., the crossover of the order parameter,  
is also applicable to MgB$_2$, one can expect that the influence of 
layered structure on MgB$_2$ should be less prominent than that on 
the highly anisotropic superconductor Bi2212. Note that besides 
the upward feature obtained, we have also achieved a 
square-root field temperature dependence \cite{Wang01b,Wang02a} 
within our framework. 

The theory reported here is generic. It can be used to study 
various properties of upper critical fields not only in different 
superconductors \cite{Wang01b}, different boundary 
conditions \cite{Wang02a,Wang02b} and different 
dimensions \cite{Wang02c}; but also possibly in 
surface \cite{James63} and $d$-wave \cite{Chang98} superconductivity and 
in the Gor'kov gap equation \cite{Gorkov59Bc2}. 
Moreover, the scheme in our theory may be used to determine 
the transition temperature of a superconductor in a magnetic field 
since the relationship between the upper critical field and 
temperature is equivalent to that between the critical temperature and 
the applied field. 

Our theory is an alternative to the traditional quantum-mechanical 
determinations of upper critical fields. The latter techniques are  
somehow related to the lowest eigenvalue (energy quantization) of 
the quantum harmonic oscillator equation. 
However, we have shown in this Letter that, instead of resorting to 
traditional method of energy quantization, the quantum harmonic 
oscillator equation can be treated just in a ``classical" manner 
(i.e., directly utilize the largest eigenvalue of the square of 
the magnetic field) to determine upper critical fields. 
It is known that some fundamental enigmas of quantum theory 
remain unresolved \cite{Tegmark01,Laloe01} and tested \cite{Leggett03}. 
Our success tempts us to ponder more about the techniques of 
quantum mechanics and to ask questions such as "what is the origin of 
the wave function (which is the central puzzle of 
quantum mechanics \cite{Tegmark01})?" 
It is clear that the concept of energy quantization can 
be bypassed in the present theory of upper critical fields. 
Moreover, the $\bm{\Psi}$ function in Eq.~(\ref{eq:Q_eigen}) is 
equivalent to the $\bm{\chi}$ function in Eq.~(\ref{eq:P_eigen}) while 
$\bm{\chi}$ is an auxiliary field to Eq.~(\ref{eq:det_P}). 
Hence $\bm{\Psi}$ is an auxiliary field? 
Whether quantum mechanics needs interpretations or not \cite{Laloe01}, 
the theory itself is of profound mathematical beauty \cite{Zeilinger00} and 
has achieved brilliant success \cite{Tegmark01}.  

L.W. thanks discussions with J.P. Ye and G.W. Wei; 
and suggestions from an anomalous referee.

\end{document}